\documentclass[12pt,preprint,amssymb]{aastex}

\usepackage{graphicx}
\usepackage[figuresright]{rotating}

\def\HI{H~{\sc i}}

\def\Ha{H$\alpha$}

\def\arcmin{$^{\prime}$}

\shorttitle{X-ray Absorption By an \HI\ Tail}
\shortauthors{Gu et al.}

\begin{document}
\title{Multi-Wavelength Studies of Spectacular Ram Pressure Stripping of a Galaxy: Discovery of an X-ray Absorption Feature}

\author {Liyi Gu\altaffilmark{1,2}, Masafumi Yagi\altaffilmark{3}, Kazuhiro Nakazawa\altaffilmark{4}, Michitoshi Yoshida\altaffilmark{5}, Yutaka Fujita\altaffilmark{6}, Takashi Hattori\altaffilmark{7}, Takuya Akahori\altaffilmark{8}, Kazuo Makishima\altaffilmark{1,4,9}}

\altaffiltext{1}{Research Center for the Early Universe, School of Science, University of Tokyo, 7-3-1, Hongo, Bunkyo-ku, Tokyo 113-0033, Japan; lygu@juno.phys.s.u-tokyo.ac.jp}
\altaffiltext{2}{Department of Physics, Shanghai Jiao Tong University, 800 Dongchuan Road, Shanghai 200240, PRC}
\altaffiltext{3}{Optical and Infrared Astronomy Division, National Astronomical Observatory of Japan, 2-21-1, Osawa, Mitaka, Tokyo, 181-8588, Japan}
\altaffiltext{4}{Department of Physics, University of Tokyo, 7-3-1, Hongo, Bunkyo-ku, Tokyo 113-0033, Japan}
\altaffiltext{5}{Hiroshima Astrophysical Science Center, Hiroshima University, 1-3-1, Kagamiyama, Higashi-Hiroshima, Hiroshima, 739-8526, Japan}
\altaffiltext{6}{Department of Earth and Space Science, Graduate School of Science, Osaka University, 1-1 Machikaneyama-cho, Toyonaka, Osaka 560-0043, Japan}
\altaffiltext{7}{Subaru Telescope, National Astronomical Observatory of Japan, 650 North A'Ohoku Place, Hilo, HI 96720, USA}
\altaffiltext{8}{Sydney Institute for Astronomy, School of Physics, The University of Sydney, NSW 2006, Australia}
\altaffiltext{9}{MAXI Team, Institute of Physical and Chemical Research, 2-1 Hirosawa, Wako, Saitama 351-0198, Japan}

%==============%
\begin{abstract}
%==============%
We report the detection of an X-ray absorption feature near the galaxy M86 in the Virgo cluster. The absorber has a column density of $2-3 \times 10^{20}$ $\rm cm^{-2}$, and its position coincides with the peak of an intracluster \HI\ cloud which was removed from the galaxy NGC 4388 presumably by ram pressure. These results indicate that the \HI\ cloud is located in front of M86 along the line-of-sight, and suggest that the stripping was primarily created by an interaction between NGC 4388 and the hot plasmas of the Virgo cluster, not the M86 halo. By calculating an X-ray temperature map, we further detected an X-ray counterpart of the \HI\ cloud up to $\approx 3^{\prime}$ south of M86. It has a temperature of 0.89~keV and a mass of $\sim 4.5$$\times 10^{8}$ $M_{\odot}$, exceeding the estimated \HI\ gas mass. The high hot-to-cold gas ratio in the cloud indicates a significant evaporation of the \HI\ gas, probably by thermal conduction from the hotter cluster plasma with a sub-Spitzer rate.

\end{abstract}
%\keywords{galaxies: clusters: general --- galaxies: evolution --- intergalactic medium --- X-rays: galaxies: clusters}

%\keywords{galaxies: clusters: individual (Abell
%  1795) --- intergalactic medium --- magnetic fields --- X-rays: galaxies: clusters}
%====================%
\section{INTRODUCTION}
%====================%

Ram pressure stripping (RPS hereafter) by intracluster medium (ICM; namely, X-ray emitting hot plasma in clusters of galaxies) is one of the major mechanisms
of gas removal from cluster member galaxies (e.g., \citealt{1972ApJ...176....1G}; \citealt{1999ApJ...516..619F}). The galactic materials, thus removed from
the galaxies, form tail-like structures behind, which have been observed in, e.g., \HI\ (e.g., \citealt{2007A&A...462...93V}), \Ha\ (e.g., \citealt{2012ApJ...749...43Y}), 
and X-ray bands (e.g., \citealt{2010ApJ...708..946S}). 
Such a gas removal process may quench the star formation in galaxies, and expedite morphological transformation from 
spirals to S0s (e.g., \citealt{1999ApJ...516..619F}; \citealt{1999ApJ...518..576P}). This provides a plausible explanation to the observed environmental effects
of cluster galaxies (e.g., \citealt{1980ApJ...236..351D}). However, details of the RPS process
are not yet clear from observational viewpoints, leaving a significant uncertainties in the current theoretical modelings.

%Note that the effects of 
%RPS in galaxy evolution and ICM enrichment rather depend on the stripping conditions; as shown in, e.g., \citet{2008MNRAS.383..593M}
%and \citet{2007A&A...466..813K}, if the RPS takes place in the cluster periphery, the resulting galaxies would be redder, and
%the ICM abundance would be higher in the outer region. However, since the line-of-sight geometries of the stripping galaxies are
%often poorly constrained, it remains rather unknown when and where does the RPS happen.   

One of the best candidates for the observational RPS study is a nearby spiral galaxy NGC 4388 (N4388 hereafter). It is located at a projected distance 
of 1.3 degree away from the core of the Virgo cluster (M87), and its 
radial velocity of $\approx 2525$ km s$^{-1}$ is about by 1500 km s$^{-1}$ higher than that of M87. Such a high velocity is thought to have
enhanced the RPS in this object. In fact, as reported in, 
e.g., \citet{1989ApJ...344..171K} and \citet{1990AJ....100..604C},
N4388 is one of the most \HI\--deficient spiral galaxies in the Virgo cluster. Recently, more intriguing RPS features have been unveiled:
\citet{2002ApJ...567..118Y} discovered in \Ha\ a 
plume of ionized gas extending 35 kpc to the northeast. Moreover, as shown in \citet[hereafter OG05]{2005A&A...437L..19O},
the stripped plume has an \HI\ counterpart with a larger extent of about 110 kpc ($\approx 24^{\prime}$), containing a mass of 
about 3.4 $\times 10^8$ $M_{\odot}$ in the atomic phase. Using an X-ray imaging analysis, \citet{2011A&A...531A..44W} found
an X-ray counterpart associated with the \Ha\ and \HI\ tails.

As suggested in, e.g, \citet{2003A&A...406..427V}, the stripping may be due to an interaction with the ICM of the Virgo cluster
at a projected distance of $\sim 350$ kpc from M87. In contrast, OG05 and \citet{2009A&A...502..427V} proposed an alternative view: 
N4388 is interacting, at a speed $> 2800$ km s$^{-1}$, with the hot gas halo of the M86 group which is assumed to be located outside the Virgo cluster. Since M86 has a projected distance of only $\sim 10$ kpc from the plume, 
they can actually be very close in three dimensions. Due to large systematic
uncertainties on the line-of-sight distance of N4388 (e.g., 13.6 Mpc from \citealt{1997A&A...323...14S}; 19.2 Mpc from \citealt{1997ApJS..109..333W}; 
and 38.7 Mpc from \citealt{2007A&A...465...71T}), the origin of the stripping in N4388
remain unclear.

The cold \HI\ plume, which has a column density sufficient to significantly absorb background X-ray photons, is found covering an X-ray bright, 
$\sim 15$\arcmin\ $\times 5$\arcmin\ field in the 
central $20$\arcmin\ region of the M86 group (OG05). This makes the X-ray absorption measurement a promising way to 
constrain the line-of-sight geometry between N4388 and M86. For this purpose, we carried out a spatially resolved
X-ray spectroscopic analysis on the M86-N4388 region. We assume a distance of 
16.7 Mpc to the Virgo Cluster, so that $1^{\prime}$ corresponds to about 4.86 kpc. For easy calculation, the same values are assumed 
also for N4388 and M86. Unless
stated otherwise, the quoted errors are at the 90\% confidence level.

%Throughout the
%paper we assume a Hubble constant of $H_0=71$$h_{71}$ km s$^{-1}$
%Mpc$^{-1}$, a flat universe with the cosmological parameters of
%$\Omega_M=0.27$ and $\Omega_\Lambda=0.73$, so that $1^{\prime}$ corresponds to about  and quote errors by the 90\%
%confidence level unless stated otherwise.

%====================%
\section{OBSERVATION AND DATA REDUCTION}
%====================%

The M86 center and N4388 regions were observed with the European Photon Imaging Camera (EPIC) onboard {\it XMM-Newton},
on 2002 July 1 and 2011 June 17, respectively. The combined field of view covers the entire 
\HI\ cloud reported in OG05. Basic reduction and calibration of the EPIC data were carried out with SAS
v12.0.1. In the screening we set {\it FLAG}=0,
and kept events with {\it PATTERNs} $0-12$ and $0-4$ for the MOS and pn cameras, respectively. By examining lightcurves extracted in
$10.0-14.0$ keV and $1.0-5.0$ keV from source-free regions, we rejected time intervals
affected by hard- and soft-band flares, respectively, in
which the count rate exceeds a 2$\sigma$ limit above the quiescent
mean value (e.g., \citealt{2004A&A...414..767K}).
The obtained clean MOS(pn) exposures are 65(39) ks and 26(10) ks for the M86 and N4388 pointings, respectively. 
Point sources detected by a SAS tool {\tt edetect\_chain} were discarded in the spectral
analysis. 
%The tool {\tt arfgen} and {\tt rmfgen} were used to
%calculate the ancillary response functions and redistribution matrices,
%respectively. 

Background was estimated as a combination of three independent components, i.e., instrumental background,
cosmic X-ray background (CXB), and the Galactic emission. Following \citet{2012ApJ...749..186G}, we created the instrumental
background spectra for all the EPIC data using a filter wheel closed dataset. The CXB and Galactic emission components were calculated 
by analyzing spectra extracted from a region near the CCD edge of the N4388 pointing, which is 
about $7^{\prime}-11^{\prime}$ south of N4388 ($25^{\prime}-30^{\prime}$ west of M87). 
Specifically, we first subtracted the instrumental background from the extracted spectra, and then fitted the 
resulting spectra with a model consisting of three components; an
absorbed power law model (photon index $\Gamma = 1.4$) describing the CXB, an absorbed optically thin thermal
model with a temperature fixed at 0.2 keV and abundance at 1.0 $Z_{\odot}$ for the Galactic foreground, and another absorbed thermal
model for the projected ICM. A same absorption column density ($N_{\rm H}$ hereafter) was applied to the three components. The best-fit temperature of the ICM component
is $1.6 \pm 0.1$ keV, which is consistent with the previous measurements (e.g., \citealt{2013MNRAS.430.2401E}).  
These fits constrained the $2.0-10.0$ keV CXB flux as 7.4 $\times 10^{-8}$ ergs cm$^{-2}$ s$^{-1}$ sr$^{-1}$, which agrees 
with other reportings (e.g., \citealt{2009PASJ...61.1117B}). Adopting the CXB brightness uncertainty of 6.2\% for a 0.5 deg$^{2}$ region
reported in \citet{2002PASJ...54..327K}, we include a CXB systematic error of 12\% in subsequent analysis. 
It was combined in quadrature with a statistical uncertainty, which was calculated by scanning over the parameter space with the 
$\tt XSPEC$ command $steppar$, to constitute the total measurement errors in the spectral fittings.

%====================%
\section{ANALYSIS AND RESULTS}
%====================%
\subsection{Excess Neutral Absorption}

In order to search the X-ray data for possible absorption features associated with the atomic gas cloud, we calculated a 2-D absorption map of the M86 central region based on a centroidal
Voronoi tessellation (CVT hereafter; \citealt{2003MNRAS.342..345C}) binning method.
The $20^{\prime}.1 \times 20^{\prime}.1$ region shown in Figure 1$a$ was thus divided into 71 sub-regions, each having a minimal signal-to-noise ratio of 100, or a count of $\geq 10000$ in $0.4-7.0$ keV. Then 
we extracted the MOS and pn spectra from individual CVT sub-regions and fit them simultaneously in $0.4-7.0$ keV 
with an absorbed APEC model for the hot gas component, plus an absorbed power-law with index of 1.6 for unresolved point sources (e.g., X-ray binaries; \citealt{2003ApJ...587..356I}) belonging to the galaxies. The 
redshift was fixed to 0, while the absorption, temperature, and metal abundance were set free. Adopting the redshift of N4388 (i.e., 0.008) does not change the fitting results. The obtained absorption map is shown in Figure 1$b$. 
The associated errors are typically
$0.8 \times 10^{20}$ $\rm cm^{-2}$ for low absorption sub-regions, and $1.0 \times 10^{20}$ $\rm cm^{-2}$ for the higher ones. We also attempted to 
apply the same spectroscopic method to the EPIC data within $6^{\prime}$ from the center of N4388; however, given the current data quality, we were not able to 
identify significant absorption structure beyond errors. Other datasets in {\it Chandra} and {\it Suzaku} archives are not suited for this purpose.

As shown in Figure 1$b$, the absorption $N_{\rm H}$ are in most regions consistent within errors with the Galactic value (i.e., $2.84 \times 10^{20}$ $\rm cm^{-2}$; \citealt{2005A&A...440..775K}) throughout the field, but we found a possible excess region $\sim 3^{\prime}$$-$$7^{\prime}$ southeast of the M86 center, as indicated by the red and purple circles. To verify this excess
absorption structure (EAS hereafter), we investigated, in details, five representative regions on and surrounding the structure. The extracted pn spectra are 
shown in Figure 2$a$. One of them (gray) is inferred to have a slightly lower temperature, as suggested by its lower spectrum peak energy than in the others. The remaining 
four exhibit similar spectra in $\geq 0.8$ keV, but two of them (red and purple) show lower fluxes at $< 0.8$ keV than the other two, suggesting different absorption $N_{\rm H}$.
To quantify these inferences, the extracted spectra were fit with the same absorbed APEC + power-law model as plotted in Figure 2$b$. The obtained confidence contours are plotted in Figure 2$c$ and 2$d$, on the absorption versus temperature, and absorption versus power-law normalization planes, respectively. It shows that the absorption on the EAS (red and purple circles; $\geq 5 \times 10^{20}$ $\rm cm^{-2}$) is higher than those of the surrounding regions 
($\sim 3 \times 10^{20}$ $\rm cm^{-2}$) at the 90\% confidence level, which cannot be ascribed to uncertainties in the gas temperature or point source measurements. As shown in \citet{2010MNRAS.403L..26C}, there is no significant Galactic dust cloud near the EAS region, indicating that the excess absorption cannot be created, either, by foreground absorption inhomogeneity.

To take into account the projected Virgo ICM component, we fit the spectra from the EAS by adding a second absorbed APEC model. According to previous measurement in \citet{2011MNRAS.414.2101U}, the temperature of the Virgo component was fixed at 2.3 keV. The two-phase model gave a better fit ($\chi^{2}/\nu = 315/279$) than the single-phase one ($\chi^{2}/\nu = 345/280$) to the northeast part of the EAS (purple circle in Fig. 1). It yielded a temperature of 0.8 $\pm$ 0.1 keV for the M86 component, and a smaller overall absorbing $N_{\rm H}$ of $3.8 \pm 0.6 \times 10^{20}$ $\rm cm^{-2}$ (Fig. 2$c$). According to the $\tt XSPEC$ tool {\tt simftest}, the fitting improvement with the two-phase model is significant at $>$ 99\% confidence level. Although the two-phase model might partially explain the observed $N_{\rm H}$ in the purple region, the excess absorption remains essentially unchanged in the red one. To further explore the origin of the EAS in the red region, we allowed the absorptions of the two-phase components to vary freely. The best-fit $N_{\rm H}$ are $5.4 \pm 1.0 \times 10^{20}$ $\rm cm^{-2}$ and $<3.7 \times 10^{20}$ $\rm cm^{-2}$ for the M86 and Virgo components, respectively. This implies that the measurement of the EAS is driven by the M86 emission. Such a two-phase model was also applied to the surrounding non-EAS regions (gray, black, and blue circles in Fig. 1), and the resulting $N_{\rm H}$ are consistent with, or smaller than, the single-phase ones.

Then we compared the detected EAS with the \HI\ tail reported in OG05. As shown in Figure 1$b$, the southwest part of the EAS nicely coincides with the peak of the \HI\ emission. The excess X-ray absorption $N_{\rm H}$ above the Galactic value, $2-3 \times 10^{20}$ $\rm cm^{-2}$, is consistent with that estimated from the radio data ($\geq 1 \times 10^{20}$ $\rm cm^{-2}$; OG05). Also, based on the two-phase results of the red and purple circles, the mass of the absorber is estimated to be $2-3 \times 10^{8} M_{\odot}$, which agrees with the \HI\ mass ($3.4 \times 10^{8} M_{\odot}$). These indicate that the southwest EAS is probably induced by the \HI\ tail. On the other hand, the northeast part of the EAS (purple circle), which exhibits an excess absorption of $\sim 1 \times 10^{20}$ $\rm cm^{-2}$ (according to the two-phase fit), appears in a more complicated region. Besides the northeast tip of the \HI\ tail, a plume of ionized gas, which has been found extending NEE towards another galaxy NGC 4438 (e.g., \citealt{2008ApJ...687L..69K}; \citealt{2008ApJ...688..208R}; \citealt{2013MNRAS.430.2401E}), also appears to overlap with this feature. Hence, the northeast part of the EAS might be caused by a superposition of different absorbers.

\subsection{A Cool X-ray tail}

To investigate thermal structure of the hot gas in the \HI\ tail, we have also created a $0.5-2.0$ keV surface brightness map and a temperature map, shown in Figure 3$a$ and 3$b$, respectively, both covering the M86 center and N4388 regions. The maps were calculated by the same CVT binning and spectral fitting methods as described in \S3.1. Figure 3$a$ reveals an X-ray emission enhancement (marked with a green box), by a factor of $\sim 3$ higher than the surrounding, extending from N4388 up to $\approx 3^{\prime}$ south of M86. It is associated apparently with the \HI\ tail. As shown in Figure 3$b$, the enhancement, or a tail, has a lower temperature, $kT$ =1.07$\pm$ 0.08 keV, than the surrounding region ($\approx 1.31$ keV) at the 99\% confidence level. 

To remove the projection effects, we carried out a two-phase spectral fitting to the low-temperature region shown in Figure 3$b$. Since the projected emission is likely to be a blend of both Virgo and M86 components, the temperature of the second APEC component was set free. This model did not give a significant improvement to the fitting ($\chi^{2}/\nu = 220/200$ compared to the single-phase $\chi^{2}/\nu = 228/202$), and the parameters of the two APEC components cannot be well constrained. This may be caused by a relatively poor data quality. We thus fixed the temperature of one component to the averaged surrounding value of 1.31 keV; then, the other temperature was obtained as 0.89$\pm$ 0.10 keV. The 1.31 keV component has a surface brightness of $\approx 3.0 \pm 0.7 \times 10^{-14}$ ergs $\rm cm^{-2}$ $\rm s^{-1}$ $\rm arcmin^{-2}$ in $0.5-7.0$ keV, consistent with the surrounding value. Hence, the 0.89 keV component is naturally considered as local emission possibly associated with the tail. Assuming a cylinder shape (45 kpc in length and 10 kpc in radius) of the extracted region shown in Figure 3$b$, the density of the 0.89 keV component was calculated to be $\approx 1.2 \pm 0.2 \times 10^{-3}$$\eta^{-0.5}$ cm$^{-3}$, where $\eta$ is the volume filling factor, and its total mass in the extracted region is $\approx 4.5 \pm 0.8 $$\times 10^{8}$$\eta^{0.5}$ $M_{\odot}$. The temperature and density of the X-ray tail are consistent with those reported in \citet{2011A&A...531A..44W}.

%The difference might be due to the large beam size (18$^{\prime\prime}\times$95$^{\prime\prime}$) 
%of the HI observation, which causes a significant underestimation of the column density.    

%====================%
\section{Summary and Discussion}
%====================%

Based on a 2-D spectral analysis using the {\it XMM-Newton} data, we detected a region with excess absorption ($\Delta$$N_{\rm H}$ = $2-3 \times 10^{20}$ $\rm cm^{-2}$) 
at a distance of $\sim 3^{\prime}$$-$$7^{\prime}$ southeast of the M86 center. The southwest peak of the absorption structure apparently coincides with the peak of an 
\HI\ gas tail which was probably removed from N4388 by ram pressure. Towards southwest of the \HI\ peak, the excess X-ray absorption diminished, but we instead observed, 
along the \HI\ tail, an excess X-ray surface brightness (by a factor of 3) and a temperature decrease (from $\sim 1.3$ keV to $0.9$ keV), which reconfirm the findings of \citet{2011A&A...531A..44W}. If this is attributed to emission from a hot plasma, we find a mass of $\sim 4.5$$\times 10^{8}$ $M_{\odot}$.

%Moreover, we detected the X-ray counterpart of the \HI\ tail, which has a
%temperature of 0.89~keV and mass of $\sim 4.5$$\times 10^{8}$ $M_{\odot}$, projected up to the center of M86 group.

%The X-ray absorption study gives direct constraint on the line-of-sight geometry between M86 and N4388. It seems to be inconsistent with some 
%of previous distance measurements, e.g., by employing methods based on Tully-Fisher and $D_{n}-\sigma$ relations for N4388 and M86, respectively, Willick et al.
%(1997) obtained a distance of 18 Mpc for N4388 and 17 Mpc for M86. It indicates that the results of nearby galaxy distance measurements should be taken
%with caveat. 

The detection of the EAS allows us to constrain, more tightly than before, the stripping location of the long \HI\ tail. Our results indicate that the EAS region of the
\HI\ tail is now in front of, or at similar distance to, the center of the M86 group. In contrast, the heliocentric recession velocity of the EAS ($\sim 2300$ km $\rm s^{-1}$; OG05) is much larger than that of 
M86 (-244 km $\rm s^{-1}$). Thus it is natural to expect that these two objects are moving towards each other at $\geq 2500$ km $\rm s^{-1}$ along the line-of-sight. Furthermore,
given an age of $\sim 200$ Myr (OG05), the EAS region of the tail should have been created at least $510$ kpc in front of M86.

This further allows us to consider the origin of the RPS. Let us first examine a possible 
interaction between N4388 and the M86 group ICM, assuming that the M86 is much outside the Virgo ICM sphere. By extrapolating the radially-averaged emission measure profile shown in Figure 9 of \citet{2008ApJ...688..208R}, 
the hot gas density of M86 
at $r=510$ kpc is $\approx 1 \times 10^{-5}$ cm$^{-3}$, and the ram pressure acted on N4388 is $\approx 1 \times 10^{-12}$ dyn cm$^{-2}$. Next we consider a 
N4388-Virgo interaction. By adopting the gas density profile reported in Figure 6 of \citet{2011MNRAS.414.2101U} and a predicted orbit of N4388 suggested in \citet{2009A&A...502..427V}, the Virgo ICM near the stripping spot ($r = 420$ kpc)
has a density of $1.5 \times 10^{-4}$ cm$^{-3}$, and the ram pressure is $\approx 4 \times 10^{-12}$ dyn cm$^{-2}$. Although the radial velocity difference between M86 and N4388 is larger than the Virgo-N4388 one, the low ICM density of the M86 significantly reduces the estimated ram pressure in the former case. A third picture is that N4388 has been interacting with
a combination of the M86 and the Virgo ICM, assuming that the M86 group is well in the Virgo cluster (\citealt{2008ApJ...688..208R}). Then the ram pressure is enhanced to $\approx 5 \times 10^{-12}$ dyn cm$^{-2}$, which is contributed mostly from the Virgo ICM. To create stripping, the ram pressure should exceed
the gravitational restoring force in the galaxy which was estimated to be $\approx 2.5 \times 10^{-12}$ dyn cm$^{-2}$ in \citet{1999ApJ...520..111V}. Hence, our current 
result suggests that the N4388-M86 interaction is insufficient for the RPS, and the \HI\ tail was created mostly by colliding with the Virgo ICM.

%In this picture we are able to derive directly the line-of-sight geometry; N4388 is located at a similar distance to M87, while M86 group is approaching from their backsides. 
%This agrees with some previous distance measurements, e.g., \citet{1993ApJ...419..479C, 2002ApJ...577...31C}, \citet{2007ApJS..169..213J}, and \citet{2010ApJ...717..603V}, 
%while it seems to disapprove some other measurements. For instance, \citet{2000A&A...355..835E} and \citet{2007A&A...465...71T} estimated the distance of N4388 to be 
%$\geq 25$ Mpc, much beyond the Virgo cluster value of $\sim 17$ Mpc, which disagrees with the EAS detection. 

Another interesting aspect is to understand phase changes of the materials removed from N4388. 
Given the associated hot plasma mass and \HI\ mass of $\sim 4.5 \pm 0.8$$\times 10^{8}$$\eta^{0.5}$ $M_{\odot}$ (\S3.2) and $3.4 \times 10^{8}$ $M_{\odot}$ (OG05), respectively, the ionized-to-atomic gas mass ratio in the tail is therefore $\sim 1$, which is an order of magnitude higher than the typical values within disk galaxies 
(e.g., \citealt{2009ApJ...697...79R}; \citealt{2012ApJ...756..183B}). It implies that the 0.89~keV plasma is not the ISM that was already in N4388 before the stripping, but rather a remnant of the cold gas, which has been transfered into the hot phase by thermal conduction or mixing with the ambient ICM. Let us examine the efficiency of the conductive heating. We may assume a minimal width of 5 kpc of the \HI\ tail (Fig. 1 of OG05) and a $N_{\rm H}$ of $2 \times 10^{20}$ $\rm cm^{-2}$ (\S 3.1). Then, employing a Spitzer rate conductivity, the evaporation timescale can be calculated to be $\sim 70$ Myr, which is a factor of 3 shorter than the age of the tail. Hence, the thermal conduction should have been suppressed to some extent. This can naturally be explained by some magnetic interfaces between the cold and hot phases. A similar scenario was proposed to explain the multi-phase hot plasmas in the cluster cD galaxy (e.g., \citealt{2001PASJ...53..401M}; \citealt{2012ApJ...749..186G}). When the conductive heat to a magnetic-field structure is substantially blocked, its \HI\ content might be condensed into star-forming materials. We will report this phenomenon in a follow-up paper. 

%Heating by disspation of turbulent motion (e.g., \citealt{2011MNRAS.414.1493R}) does not work efficiently either. 

\section*{Acknowledgments}
We thank Tom Oosterloo and Jacqueline van Gorkom for providing us the \HI\ map.

\begin{figure}
\begin{center}
\includegraphics[angle=-0,scale=.4]{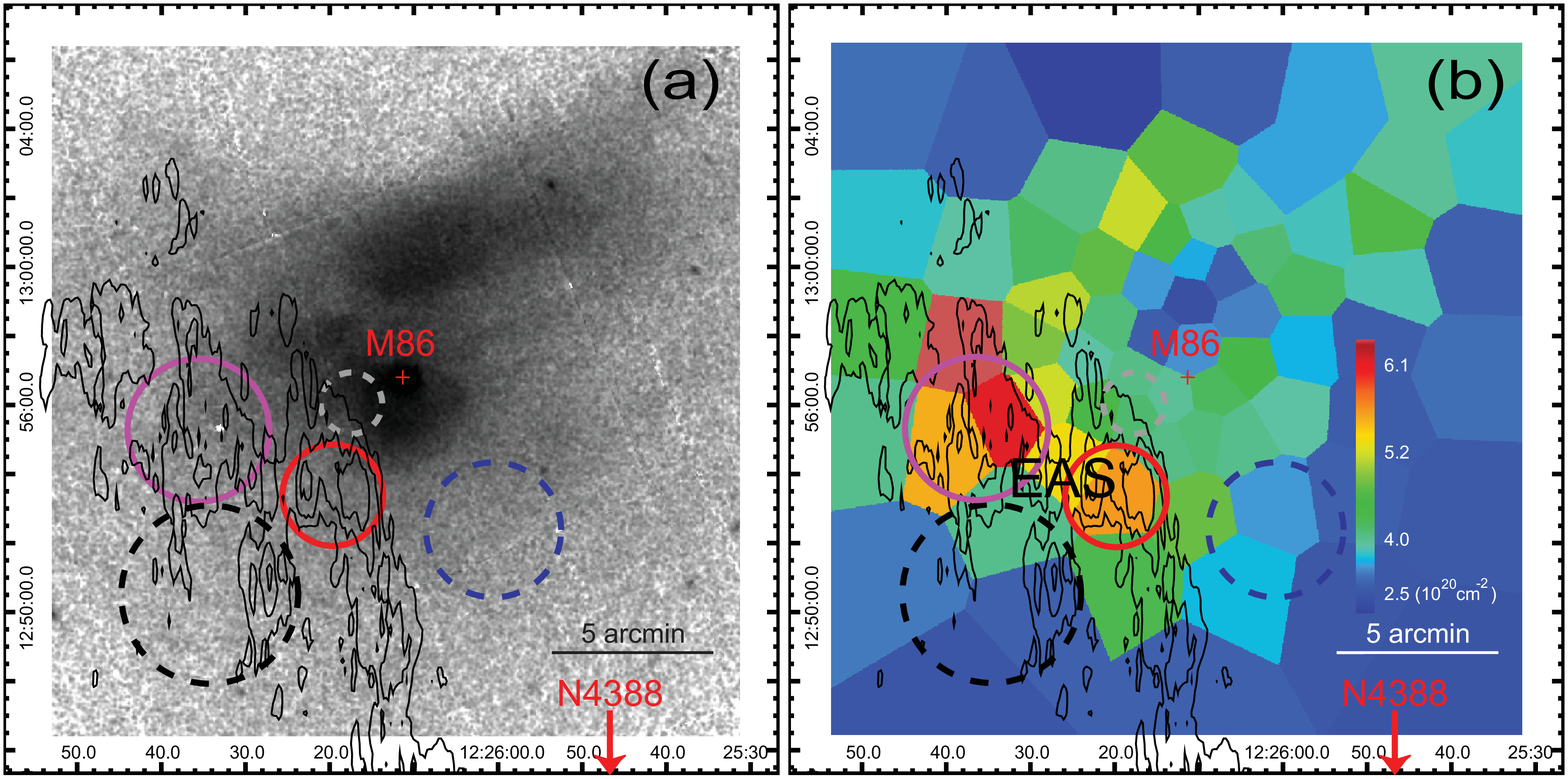}
\caption{($a$) A $0.5-2.0$ keV exposure-corrected image of the M86 group. The \HI\ brightness contours are overlaid in black, with contour levels of 1.0, 5.0, and 10.0 in units of $10^{19}$ cm$^{-2}$. The regions for detailed spectral analysis (\S3.1) are shown with circles, where red and purple solid circles indicate southwest and northeast of the EAS, respectively, and gray, black, and blue dashed ones show the surrounding regions. ($b$) X-ray absorption map of the M86 central region. Errors are typically
$0.8 \times 10^{20}$ $\rm cm^{-2}$ for low absorption regions, and $1.0 \times 10^{20}$ $\rm cm^{-2}$ for higher ones.}
\end{center}
\end{figure}

\begin{figure}
\begin{center}
\includegraphics[angle=-0,scale=.6]{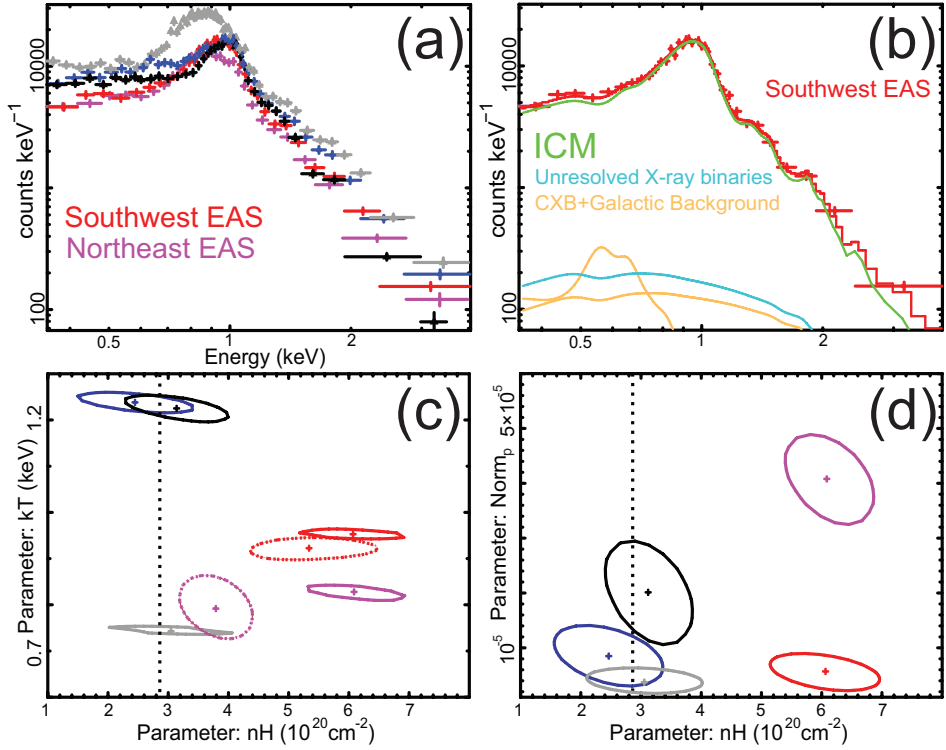}
\caption{($a$) EPIC-pn spectra of the five regions shown in Figure 1. The spectra are colored in the same way as the circles in Figure 1. ($b$) Best-fit single-phase models to the EPIC-pn spectrum of the southwest part of the EAS. The ICM, unresolved point source, and background components are shown in green, cyan, and orange, respectively. ($c$) Confidence contours between temperature and absorption at the 90\% confidence level for the five regions. Solid and dotted contours indicate the results with single-phase and two-phase models. The temperatures of the cooler components are plotted in the two-phase cases. The Galactic \HI\ column density is indicated by a dashed black line. ($d$) Confidence contours between the normalization of unresolved point source component and absorption obtained with the single-phase model at the 90\% confidence level.}
\end{center}
\end{figure}

\begin{figure}
\begin{center}
\includegraphics[angle=-0,scale=.5]{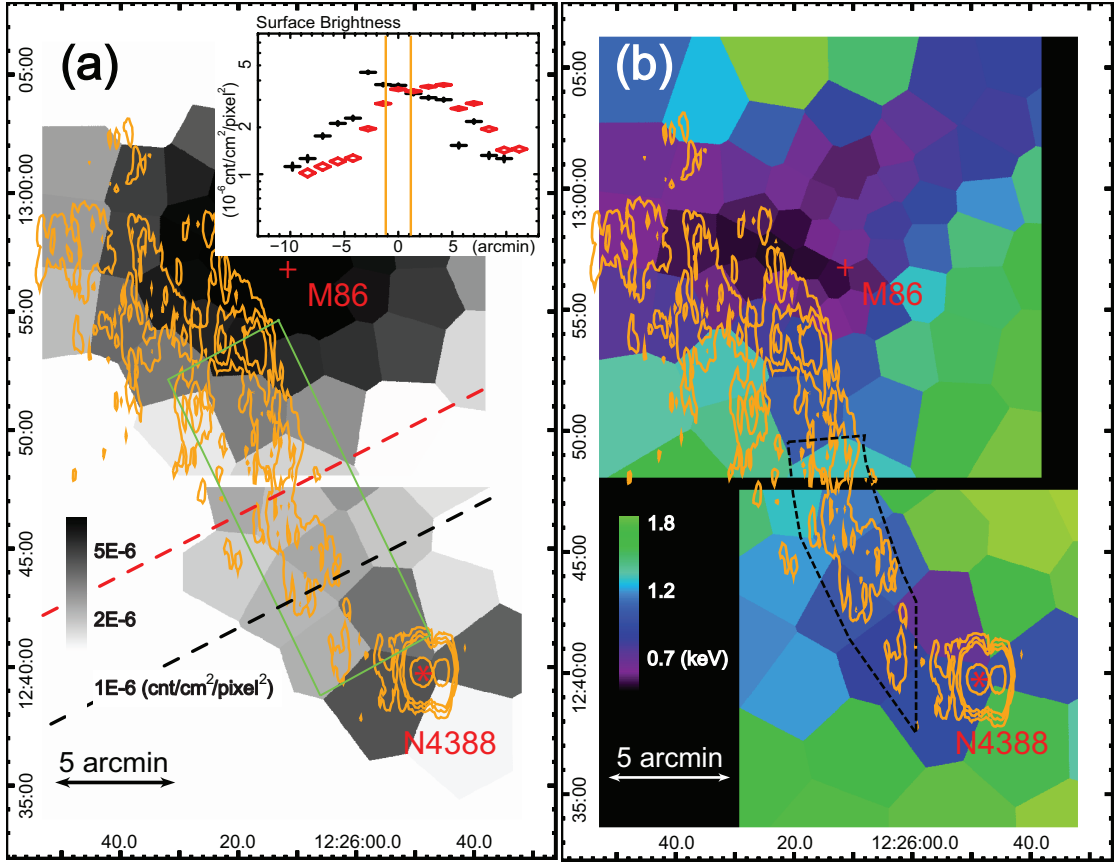}
\caption{($a$) A $0.5-1.0$ keV exposure-corrected, background-subtracted surface brightness map of the M86-N4388 region. The \HI\ brightness contours are overlaid in orange, and the centers of M86 and N4388 are shown with a red 
cross and a star, respectively. A green box indicate the region with enhanced X-ray brightness. Surface brightness profiles along the dashed red and black lines are shown in the insert, where the \HI\ cloud is marked with vertical orange lines. $1\times 10^{-6}$ cnt cm$^{-2}$ pixel$^{-2}$ is about $2.7 \times 10^{-14}$ ergs cm$^{-2}$ s$^{-1}$ arcmin$^{-2}$ in $0.5-7.0$ keV. ($b$) A temperature map of the same region. The region used for detailed spectral analysis (\S3.2) is marked with dashed black lines.}
\end{center}
\end{figure}

\end{document}